# A Leakage-Aware Comparative Benchmark of Machine Learning, Deep Learning, and Transformer Models for Reliable Leukemia Detection

**Nisreen Albzour**

*School of Systems Science and Industrial Engineering, Binghamton University, Binghamton, NY, USA*

ORCID: https://orcid.org/0009-0000-9317-340X

## Abstract

Automated classification of acute lymphoblastic leukemia (ALL) from peripheral blood smear images has often reported near-perfect performance on the C-NMC 2019 dataset. We show that such results may be inflated by patient-level data leakage associated with random image-level partitioning, where cells from the same subject may appear in both training and test folds. We establish a leakage-aware benchmark under a strict subject-disjoint protocol, comparing LightGBM, RBF-SVM, EfficientNet-B0, EfficientNet-B1, and ViT-Tiny. Models are developed using three subject-disjoint folds from 73 subjects and evaluated on an external preliminary-phase test set of 1,867 images from 28 unseen subjects with zero patient overlap. Beyond discrimination, we assess calibration using expected calibration error, Brier score, and temperature scaling. Under honest evaluation and averaged across three training seeds, EfficientNet-B0 achieves the best and most consistent performance, with AUROC 0.882 ± 0.016, sensitivity 0.899 ± 0.028, specificity 0.666 ± 0.007, and calibrated ECE 0.033 ± 0.012; EfficientNet-B1 attains comparable but more variable performance (AUROC 0.838 ± 0.028). Frozen-feature classifiers and ViT-Tiny show high sensitivity but poor specificity, indicating a tendency to over-predict the malignant class. A random-versus-subject-disjoint ablation shows that random splitting inflates AUROC by about 0.04 even in the conservative frozen-feature setting. These findings caution against image-level evaluation on C-NMC 2019 and provide a reproducible, calibration-aware benchmark for future work.

## 1. Introduction

Acute lymphoblastic leukemia (ALL) is an aggressive hematologic malignancy characterized by the abnormal proliferation of immature lymphoid cells. Accurate and timely identification is important because diagnostic delay can affect treatment planning and clinical outcomes. In routine practice, the diagnostic pathway commonly begins with clinical assessment and laboratory testing, including complete blood count and microscopic examination of peripheral blood or bone marrow smears, followed by confirmatory immunophenotypic, cytogenetic, and molecular analyses when indicated. Although definitive diagnosis requires comprehensive laboratory evaluation, microscopic cell morphology remains an important component of hematopathology workflows and provides a natural target for computer-assisted image analysis [1]. As shown in Figure 1, ALL is one of the major leukemia subtypes arising from abnormal white blood cell morphology, distinguishing it from acute myeloid leukemia (AML), chronic lymphocytic leukemia (CLL), and chronic myeloid leukemia (CML).

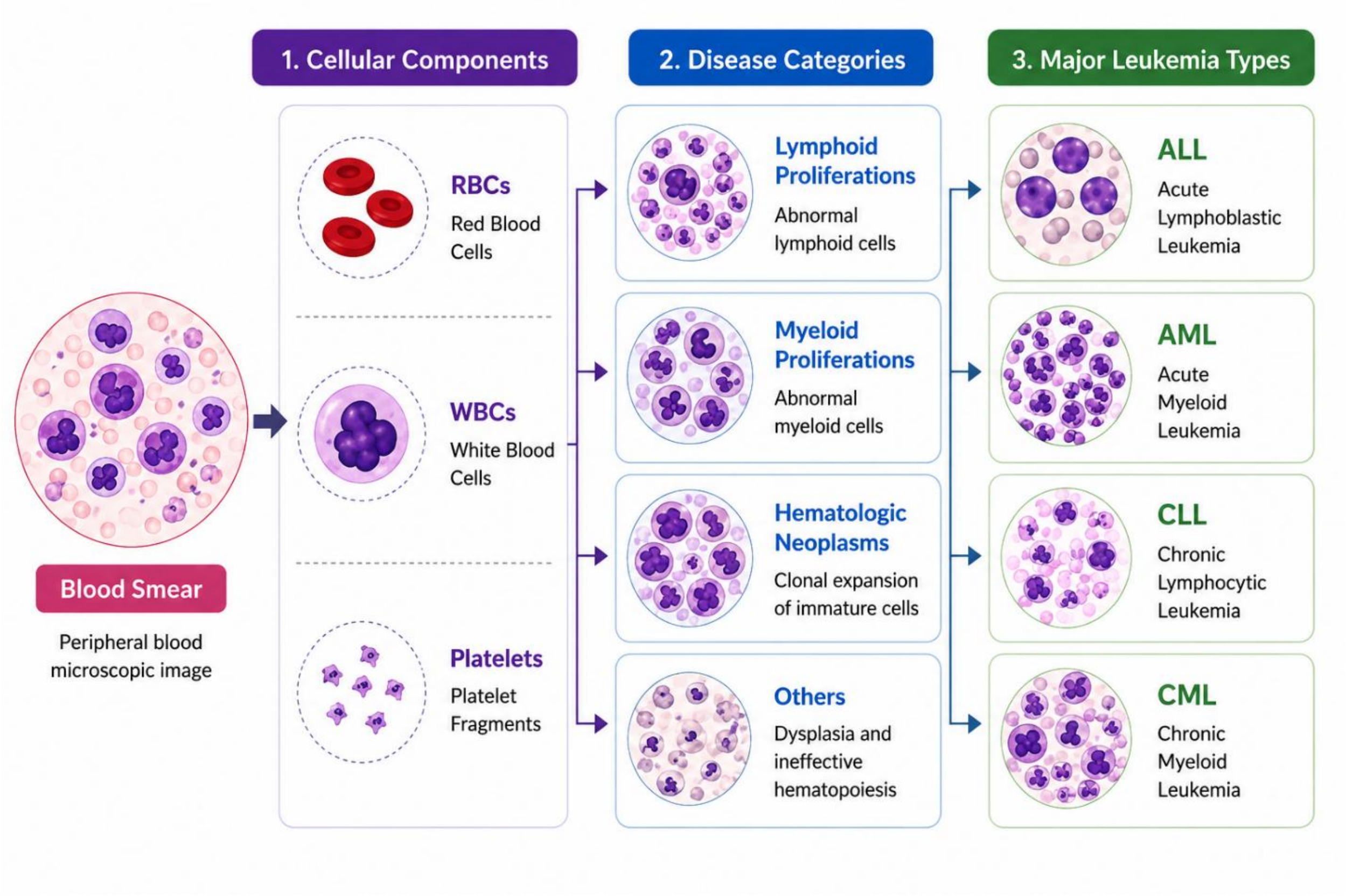

*Figure 1. From peripheral blood smear to leukemia subtype classification. A blood smear yields red blood cells (RBCs), white blood cells (WBCs), and platelets; abnormal WBC morphology is grouped into disease categories (lymphoid proliferations, myeloid proliferations, hematologic neoplasms, and other dysplastic changes), which in turn encompass the major leukemia types: acute lymphoblastic leukemia (ALL), acute myeloid leukemia (AML), chronic lymphocytic leukemia (CLL), and chronic myeloid leukemia (CML). This study focuses on ALL classification.*

Manual microscopic examination is labor-intensive, requires experienced personnel, and may be affected by inter-observer variability, staining differences, and workload. These challenges have motivated the development of automated leukemia screening systems based on machine learning and deep learning. In recent years, convolutional neural networks, transfer learning, ensemble methods, and transformer-based models have been applied to blood-smear and cancer-cell images, often reporting very high classification performance [2], [3], [4], [5]. Such systems are promising as decision-support tools, particularly for triage, workload reduction, and settings where expert hematopathology resources are limited. However, high internal performance does not necessarily imply reliable generalization to unseen patients.

The C-NMC 2019 dataset, released for the ISBI 2019 challenge, has become one of the most widely used public benchmarks for ALL cell classification [6], [7]. It contains labeled single-cell images organized by subject-level folds, together with a labeled preliminary-phase set and an unlabeled final-phase set [2]. Its public availability has enabled extensive comparison of classical machine learning, convolutional neural networks, automated machine learning, and more recently transformer-based approaches [3], [4], [5]. Nevertheless, the dataset also contains multiple cell images per subject, making the evaluation protocol especially important. If images are split randomly at the image level, cells from the same subject may appear in both training and test sets. In that setting, a model can exploit subject-specific staining, acquisition, or morphology patterns [8] rather than learning disease-relevant features, which may contribute to inflated performance estimates.

This issue is a specific form of data leakage, a well-recognized threat to reproducibility in machine-learning-based science [9]. In medical imaging, leakage can be particularly subtle because observations from the same patient are often highly correlated. A random image-level split may therefore evaluate recognition of patients or acquisition conditions rather than generalization to new patients. For C-NMC 2019, where multiple cells are available per subject, subject-disjoint evaluation is essential: all images from a given subject must remain within a single fold or test partition. Without this safeguard, near-perfect reported scores may overstate clinical utility and obscure the true difficulty of the task. This concern is consistent with broader evidence from cancer-prediction studies showing that models that perform well internally may degrade on independent cohorts [10].

Beyond leakage-free discrimination, reliability is also important. A leukemia screening model should not only rank malignant and normal cells correctly, but also provide probability estimates that are meaningful. Poorly calibrated models may be overconfident in incorrect predictions, which is problematic for clinical decision support. Metrics such as expected calibration error and Brier score, together with post-hoc methods such as temperature scaling, provide complementary information to AUROC, accuracy, and F1-score [11]. Similar calibration-aware, imbalance-sensitive ensemble studies in cervical cytology have shown that reliability gains can be achieved without sacrificing discrimination [12]. Despite their importance, calibration and reliability remain underreported in leukemia cell classification studies, where headline discrimination metrics are often emphasized over probability trustworthiness.

In this work, we establish a leakage-aware and calibration-aware benchmark on C-NMC 2019. We compare classical machine learning models trained on frozen EfficientNet-B0 features [13], [14], fine-tuned EfficientNet models at two scales [15], and a compact vision transformer [16] under the same strict subject-disjoint protocol. Model development uses the three predefined subject-disjoint training folds, while final evaluation is performed on the labeled preliminary-phase set as an external held-out test set with zero subject overlap. We further quantify leakage-induced inflation by comparing random image-level splitting against subject-disjoint splitting for the classical models.

Our contribution is not a new architecture, but, to our knowledge, the first benchmark on C-NMC 2019 to jointly combine strict subject-disjoint evaluation, a genuinely external test set, calibration analysis, and statistically validated model comparison—clarifying how evaluation protocol, model family, scaling, and calibration affect reported performance on this dataset. Specifically, we make four contributions. First, we establish a subject-disjoint evaluation protocol that prevents patient-level leakage and uses a separate preliminary-phase set as a genuinely external test with zero subject overlap. Second, we compare frozen-feature machine learning, fine-tuned convolutional networks, and a compact vision transformer under identical, leakage-free conditions. Third, we move beyond discrimination by reporting calibration metrics, temperature scaling behavior, and confusion structure, together with statistically validated pairwise model comparisons. Fourth, we directly quantify leakage-induced AUROC inflation via a controlled random-versus-subject-disjoint ablation.

## 2. Related Work

This section reviews the literature most relevant to the present work, covering deep learning and classical machine learning approaches to ALL classification, the use of frozen deep features with lightweight classifiers, vision transformers in medical imaging, evaluation protocol and data leakage, and model calibration in clinical decision support. Figure 2 summarizes these six areas.

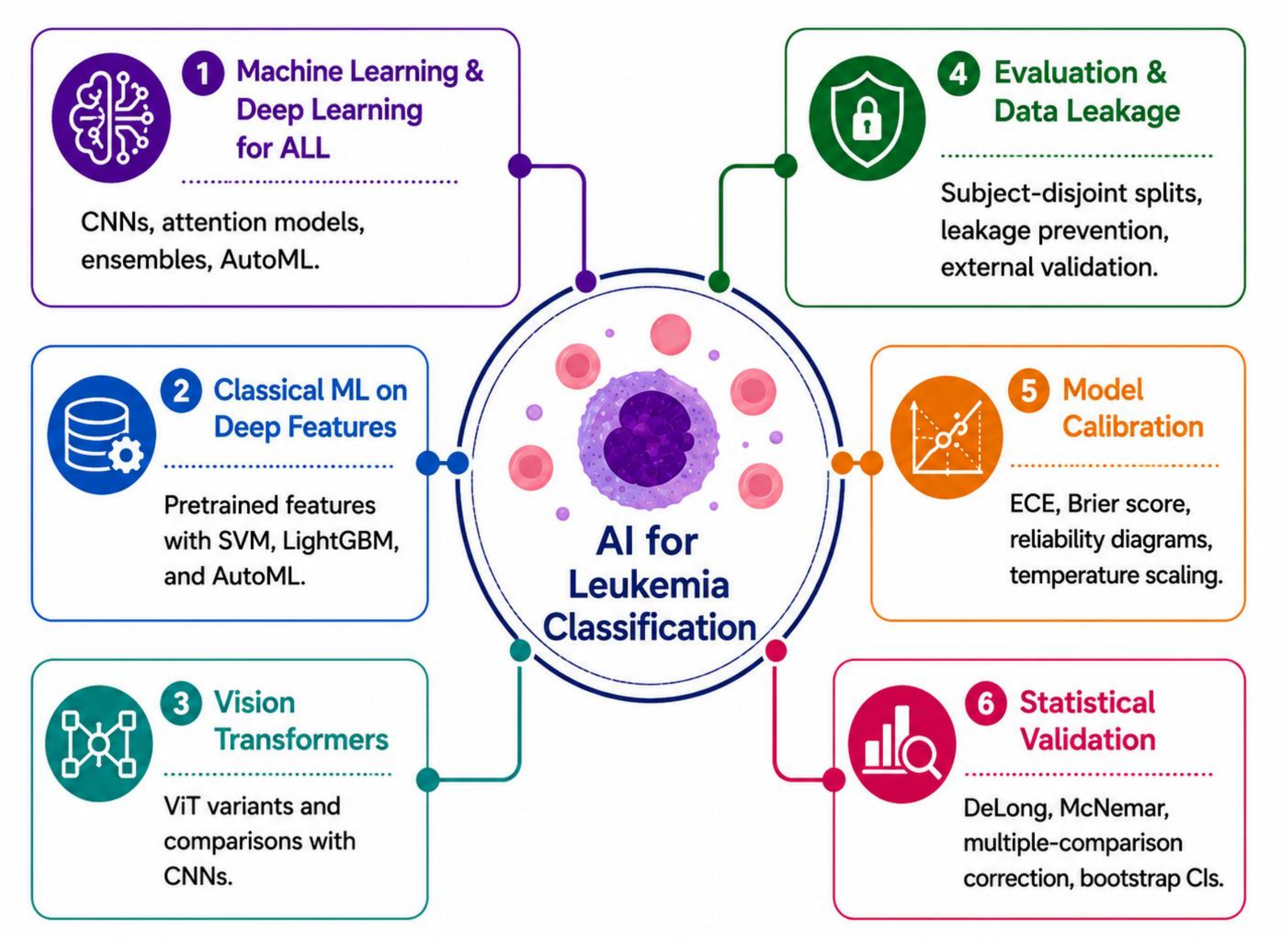

*Figure 2. Overview of AI techniques applied to leukemia detection, spanning machine learning and deep learning approaches, classical machine learning on deep features, vision transformers, evaluation protocol and data leakage, model calibration, and statistical validation.*

## 2.1 Machine Learning and Deep Learning for ALL Classification

Automated classification of acute lymphoblastic leukemia (ALL) cells has been pursued with a wide range of methods. Challenge submissions to the ISBI 2019 C-NMC task reported F1-scores in the high eighties to low nineties on the held-out test phase [5], with subsequent work pushing further using transfer-learning backbones such as DenseNet [17], EfficientNet [15], and Inception variants. Mondal et al. [4] developed a weighted ensemble of pre-trained CNNs (VGG-16, Xception, MobileNet, InceptionResNet-V2, DenseNet-121), achieving 86.2% average accuracy on C-NMC 2019. More recently, attention-augmented architectures have combined EfficientNet with squeeze-and-excitation mechanisms, reporting F1-scores above 97% on the C-NMC 2019 benchmark [18], while hybrid pipelines combining ResNet deep feature extraction with metaheuristic feature selection and an MLP classifier have achieved roughly 90.7% accuracy on the same dataset [19]. Neural ensemble methods further combining memetic optimization with deep feature selection have also been proposed for ALL detection, demonstrating that carefully tuned feature selection can meaningfully boost discrimination over single-model baselines [20]. Automated machine-learning frameworks have similarly been applied to leukemia classification, enabling systematic search across model families without manual tuning [3]. A consistent theme across this literature is that near-perfect performance is frequently reported, yet the evaluation protocols underpinning these results vary substantially and are often inadequately described. The absence of standardized, subject-disjoint evaluation makes it difficult to determine whether reported gains reflect genuine diagnostic capability or artifacts of the partitioning strategy. To our knowledge, no prior study on C-NMC 2019 has combined strict subject-disjoint evaluation, a genuinely external test set, calibration analysis, and statistically validated model comparison within a single benchmark, a combination that is central to establishing whether reported performance reflects diagnostic capability or evaluation artifacts. Beyond these, attention-augmented and ensemble approaches have continued to report strong results on ALL classification [21], [22], [23], [24], building on earlier demonstrations of deep learning for this task, such as the segmentation-plus-CNN approach of Rehman et al. [25], which established the feasibility of automated diagnosis on blood-smear images. Systematic reviews of this literature corroborate both the rapid methodological progress and the persistent heterogeneity in datasets and evaluation protocols [26].

### 2.2 Classical Machine Learning on Deep Features

Beyond end-to-end training, frozen deep features extracted from ImageNet-pretrained networks have served as input to classical classifiers including support vector machines [13] and gradient-boosted trees such as LightGBM [14]. Al-Zoubi and Al-Bzoor [3] demonstrated competitive performance on C-NMC 2019 using automated machine-learning pipelines for leukemia detection. These frozen-feature approaches are valuable as diagnostic baselines because, unlike fine-tuned networks, they have limited capacity to memorize subject-specific texture during training, making them a conservative probe for quantifying leakage-induced performance inflation. Support vector machines with radial-basis-function kernels have been widely used in cytology image analysis owing to their strong generalization in high-dimensional feature spaces and their well-understood margin-based decision boundaries [13]. LightGBM, a histogram-based gradient-boosting framework, offers efficient training on large feature vectors with built-in support for class imbalance through sample weighting, directly relevant to the approximately 2:1 ALL-to-normal imbalance present in C-NMC 2019 [14]. The combination of powerful frozen deep representations with lightweight classical classifiers also provides a practical deployment pathway in resource-constrained clinical settings where re-training deep networks is not feasible. Classical models such as SVM, decision trees, random forest, and AdaBoost remain competitive baselines more broadly in clinical outcome prediction, particularly when combined with feature selection or dimensionality reduction on structured assessment data [27].

### 2.3 Vision Transformers in Medical Imaging

Vision transformers (ViTs) [16], originally designed for large-scale natural-image recognition, have been applied increasingly to cytology and histology classification. Comparative studies consistently find that compact transformers underperform fine-tuned CNNs on small or moderately sized medical datasets, where their weaker inductive spatial bias and data-hungry self-attention require substantially more training samples to converge effectively [28], [29], [30]. Albzour and Lam [28], [29] benchmarked ViT-Tiny against CNN baselines for cervical-cell classification, finding that fine-tuned CNNs remain superior under strict subject-disjoint evaluation with statistical validation. A recent review of vision transformers in medical imaging [30] similarly notes that context-rich tasks with abundant data favor transformers, while data-scarce regimes with local morphological discriminators—precisely the setting of leukemia cell classification—continue to favor CNNs. The interpretability of transformer attention maps has been cited as a potential clinical advantage, as attention-rollout and Grad-CAM-equivalent visualizations can highlight morphologically relevant regions of a cell; however, this must be weighed against the higher data requirements and computational cost of transformer architectures relative to compact CNNs [30]. Hybrid approaches that incorporate transformer modules within CNN architectures have been proposed as a middle ground, aiming to combine the local feature sensitivity of CNNs with the global context modeling of transformers, though rigorous subject-disjoint benchmarks comparing such hybrids on C-NMC 2019 remain scarce.

### 2.4 Data Leakage and Evaluation Protocol

Data leakage is a well-recognized threat to reproducibility in machine-learning-based science [9]. In medical imaging, the most common form arises from image-level rather than patient-level data splitting: when multiple images per subject are available, random splitting places cells from the same patient into both training and test partitions, enabling models to exploit subject-specific staining or acquisition artifacts rather than disease-relevant morphology [8]. Studies of longitudinal brain MRI demonstrate that subject-wise splitting is essential to prevent identity confounding, whereby a model learns to identify subjects alongside diagnostic features [31]. For C-NMC 2019 specifically, the inconsistency between the modest F1-scores of the original challenge submissions [5], [7] and the near-perfect scores later reported by image-level evaluations may be partly explained by patient-level leakage associated with image-level evaluation. Our subject-disjoint ablation directly quantifies this inflation for the frozen-feature models tested here.

### 2.5 Model Calibration in Clinical Decision Support

Beyond discrimination, reliable clinical deployment requires probability estimates that are meaningful and trustworthy. Poorly calibrated models may be overconfident in incorrect predictions, which is harmful when probabilities are used to triage patients or communicate diagnostic uncertainty. Guo et al. [11] established temperature scaling as a simple and effective post-hoc recalibration method, demonstrating that modern deep networks are systematically overconfident. However, cross-cohort studies have found that calibration and discrimination can decouple under distribution shift—a model well calibrated on development data may not remain calibrated on an independent cohort [10], [32], [33]. Despite this, calibration is routinely omitted from leukemia cell classification studies, which predominantly report accuracy, F1-score, and AUROC as headline metrics. Our benchmark explicitly addresses this gap by reporting ECE and Brier score before and after temperature scaling on a genuinely external, subject-disjoint test set. Broader reviews of uncertainty quantification in medical image analysis similarly stress calibration as a core component of trustworthy clinical AI [33].

These reliability challenges are not unique to medical imaging: in other high-class-imbalance classification settings, such as fine-grained emotion classification, imbalance-handling strategies and calibration-aware selective prediction have similarly been shown to affect real-world reliability beyond what raw accuracy reports [34], [35], reinforcing the case for reporting calibration alongside discrimination in the present benchmark. Reliability-aware quality-control frameworks have similarly been proposed outside classification tasks, for instance converting multimodal image registration quality into interpretable risk categories using data-learned thresholds [36], underscoring that per-case reliability reporting is a broader need across medical imaging pipelines.

### 2.6 Statistical Evaluation and Significance Testing

A growing body of work argues that performance comparisons in medical-imaging AI are frequently reported without statistical support. A recent systematic analysis of classification and segmentation papers found that, although most reported the superiority of a proposed method, only a small minority justified their claims with significance testing, and a substantial fraction of the outperformance claims had a high probability of arising by chance [37]. Related critiques of biomedical-image competitions and of common evaluation practice have reached similar conclusions, noting that apparent ranking differences often lie within the range of statistical noise [38], [39]. These findings motivate the explicit statistical comparison we report in Section 4.7.

Established methodology exists for such comparisons. Demšar [40] provides widely used recommendations for comparing classifiers, and Dietterich [41] characterizes the trade-offs of approximate statistical tests for supervised learning. For discrimination specifically, the DeLong test [42] compares the areas under two correlated ROC curves, while McNemar's test [43] assesses paired differences in classification outcomes. When several comparisons are made, the Holm correction [44] controls the family-wise error rate without the conservativeness of a simple Bonferroni adjustment. Confidence intervals for performance metrics are commonly obtained by bootstrap resampling [45]; in subject-structured data, resampling at the patient level rather than the image level is necessary to avoid underestimating uncertainty. We adopt these methods, reporting DeLong and McNemar tests with Holm correction and patient-level bootstrap confidence intervals.

# 3. Materials and Methods

Figure 3 provides an overview of the complete leakage-aware benchmarking pipeline, from dataset preparation through subject-disjoint evaluation, model training, calibration analysis, leakage ablation, and final performance and interpretability assessment. Each stage is detailed in the subsections that follow.

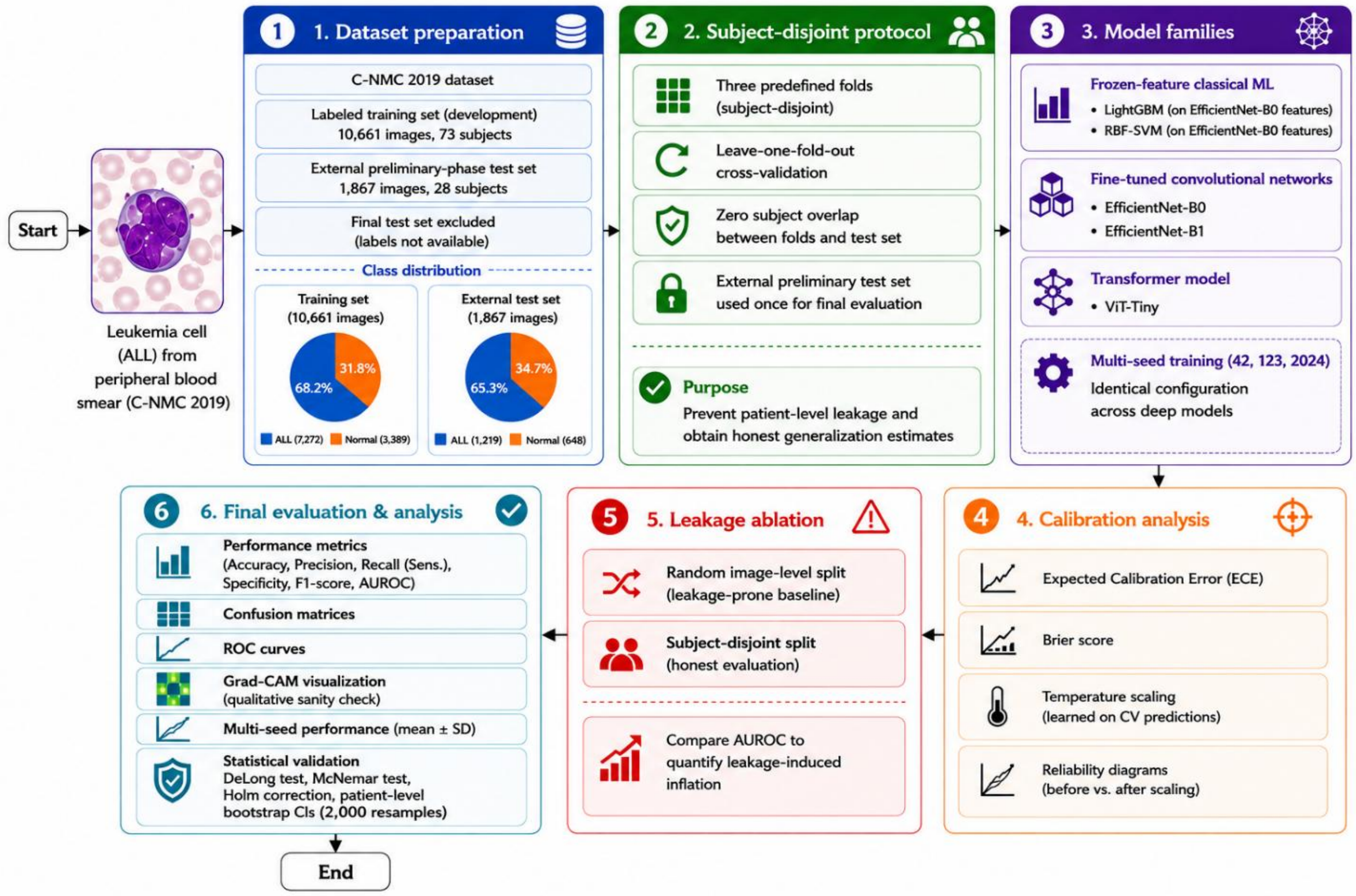

*Figure 3. Overview of the leakage-aware benchmarking workflow for C-NMC 2019. The pipeline begins with leukemia cell images, proceeds through dataset preparation, subject-disjoint evaluation, multiple model families, calibration analysis, and leakage ablation, and concludes with comprehensive performance evaluation and interpretability assessment.*

## 3.1 Dataset

We use the C-NMC 2019 dataset [6], [7]. The labeled training data comprise 10,661 single-cell images (7,272 ALL, 3,389 normal) from 73 subjects, organized into three folds. The labeled preliminary-phase set contains 1,867 images (1,219 ALL, 648 normal) from 28 subjects. The official final-phase set (2,586 images) is excluded throughout because its labels are not publicly available. We verified that the three training folds are mutually subject-disjoint and that no subject in the preliminary set appears in training. As shown in Figure 4, both partitions are imbalanced toward the malignant (ALL) class at an approximately 2:1 ratio, motivating the use of class-balanced training and the reporting of specificity alongside sensitivity.

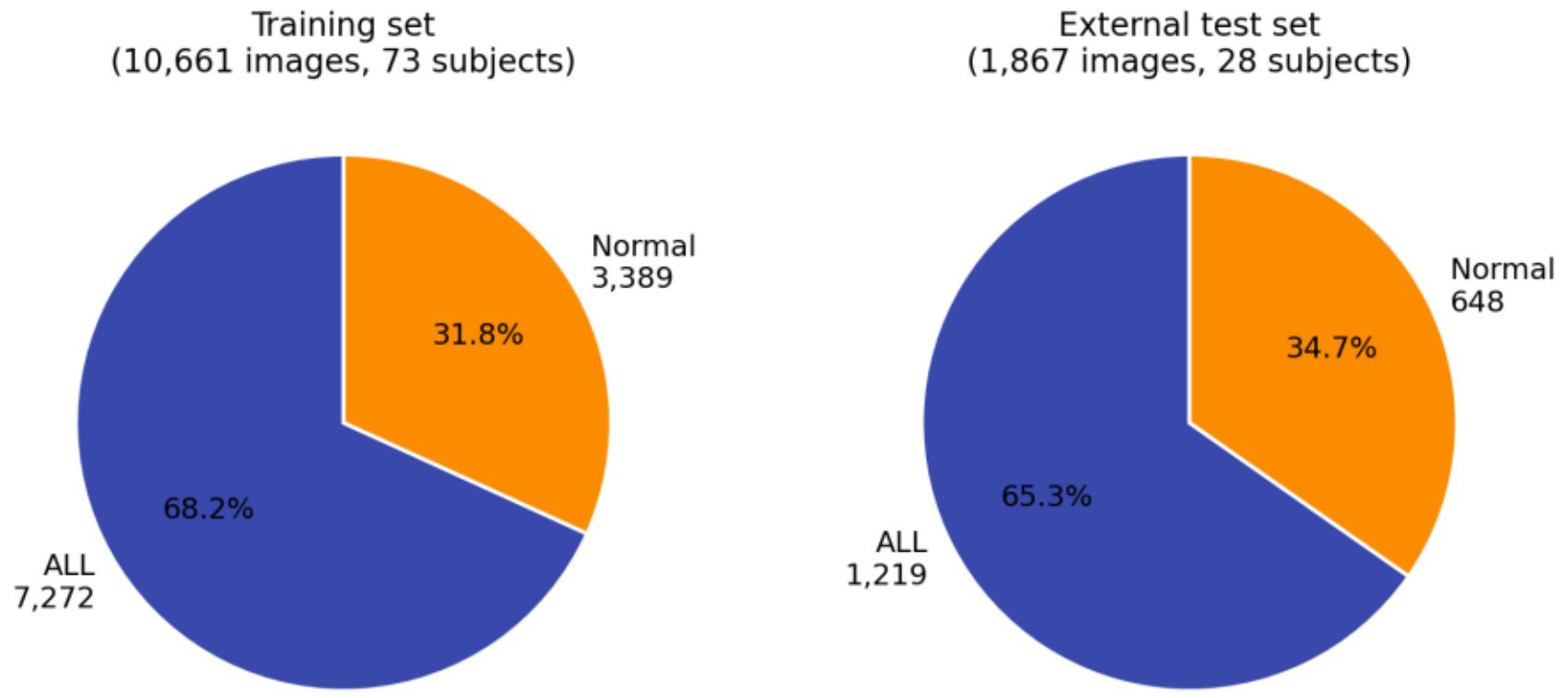

*Figure 4. Class distribution of the C-NMC 2019 partitions used in this study: the training set (left) and the external preliminary-phase test set (right). Both are imbalanced toward ALL at roughly 2:1.*

### 3.2 Evaluation protocol

All evaluation is subject-disjoint, with the deliberate exception of the image-level random split used in the leakage ablation (Sections 3.5 and 4.4), in line with reproducible benchmarking practice that emphasizes statistically valid, leakage-free comparison [29], [30]. Model development uses three-fold leave-one-fold-out cross-validation (LOFO-CV) across the 73 training subjects. Final performance is reported on the preliminary-phase set as a genuinely external test of 28 unseen subjects. Crucially, no cell from a given patient ever appears in more than one partition. Hyperparameter and calibration decisions are made only on cross-validation folds; the external test set is touched once. For the deep models, external predictions are the mean of the three cross-validation models, so that all folds contribute.

### 3.3 Models and hyperparameters

We evaluate five models spanning three families: two classical machine-learning classifiers on frozen deep features, two fine-tuned convolutional networks, and one vision transformer. Table 1 lists the full roster. The two classical models are configured as in Table 2; the three deep models share the identical training configuration in Table 3, differing only in backbone architecture, ImageNet-pretrained initialization, parameter count, and classification head, so that the comparison isolates the effect of architecture rather than tuning. All deep models are optimized with AdamW [46]. Figure 5 illustrates the architecture of the fine-tuned EfficientNet-B1 model, representative of the EfficientNet family evaluated here, including its compound-scaled backbone and the dropout-plus-linear classification head used for the EfficientNet models.

*Table 1. The five models evaluated in this study, grouped by family. The classical models use features extracted from a frozen EfficientNet-B0; the deep models are fine-tuned end-to-end.*

| Model | Family | Feature source / backbone | Training |
|---|---|---|---|
| LightGBM | Gradient boosting | Frozen EfficientNet-B0 features | Table 2 |
| RBF-SVM | Kernel SVM | Frozen EfficientNet-B0 features | Table 2 |
| EfficientNet-B0 | CNN (fine-tuned) | EfficientNet-B0 backbone | Table 3 |
| EfficientNet-B1 | CNN (fine-tuned) | EfficientNet-B1 backbone | Table 3 |
| ViT-Tiny | Vision transformer | ViT-Tiny backbone | Table 3 |

*Table 2. Hyperparameters of the classical machine-learning models trained on frozen EfficientNet-B0 features.*

| Hyperparameter | LightGBM | RBF-SVM |
|---|---|---|
| Input features | Frozen EfficientNet-B0 (1,280-d) | Frozen EfficientNet-B0 (1,280-d) |
| Estimators / kernel | n_estimators = 400 | kernel = RBF |
| Learning rate | 0.05 | — |
| Tree / margin param. | num_leaves = 31 | C = 1.0 |
| Kernel coefficient | — | gamma = scale |
| Subsampling | subsample = 0.8 | — |
| Feature subsampling | colsample_bytree = 0.8 | — |
| Feature scaling | — | StandardScaler |
| Class imbalance | class_weight = balanced | class_weight = balanced |
| Probability estimates | native | Platt scaling |
| Random state | 42 | 42 |

*Table 3. Shared training configuration for the deep models. The models differ only in backbone (EfficientNet-B0, EfficientNet-B1, ViT-Tiny), pretrained initialization, parameter count, and classification head.*

| Setting | Value (shared by EfficientNet-B0, EfficientNet-B1, ViT-Tiny) |
|---|---|
| Initialization | ImageNet-pretrained weights |
| Input resolution | 224 × 224 RGB |
| Optimizer | AdamW |
| Learning rate | $3 \times 10^{-4}$ |
| Weight decay | $1 \times 10^{-4}$ |
| Batch size | 32 |
| Max epochs | 8 (early stopping, patience = 3 on held-out fold AUROC) |
| Loss | BCE-with-logits, class-balanced positive weight (≈ 0.47) |
| Precision | Mixed precision (AMP) |
| Augmentation | Random H/V flip, rotation (±15°), mild color jitter |
| Classification head | Dropout (0.3) + single linear logit (EfficientNet-B0/B1); single linear logit (ViT-Tiny) |
| External prediction | Mean of the three LOFO-CV models |

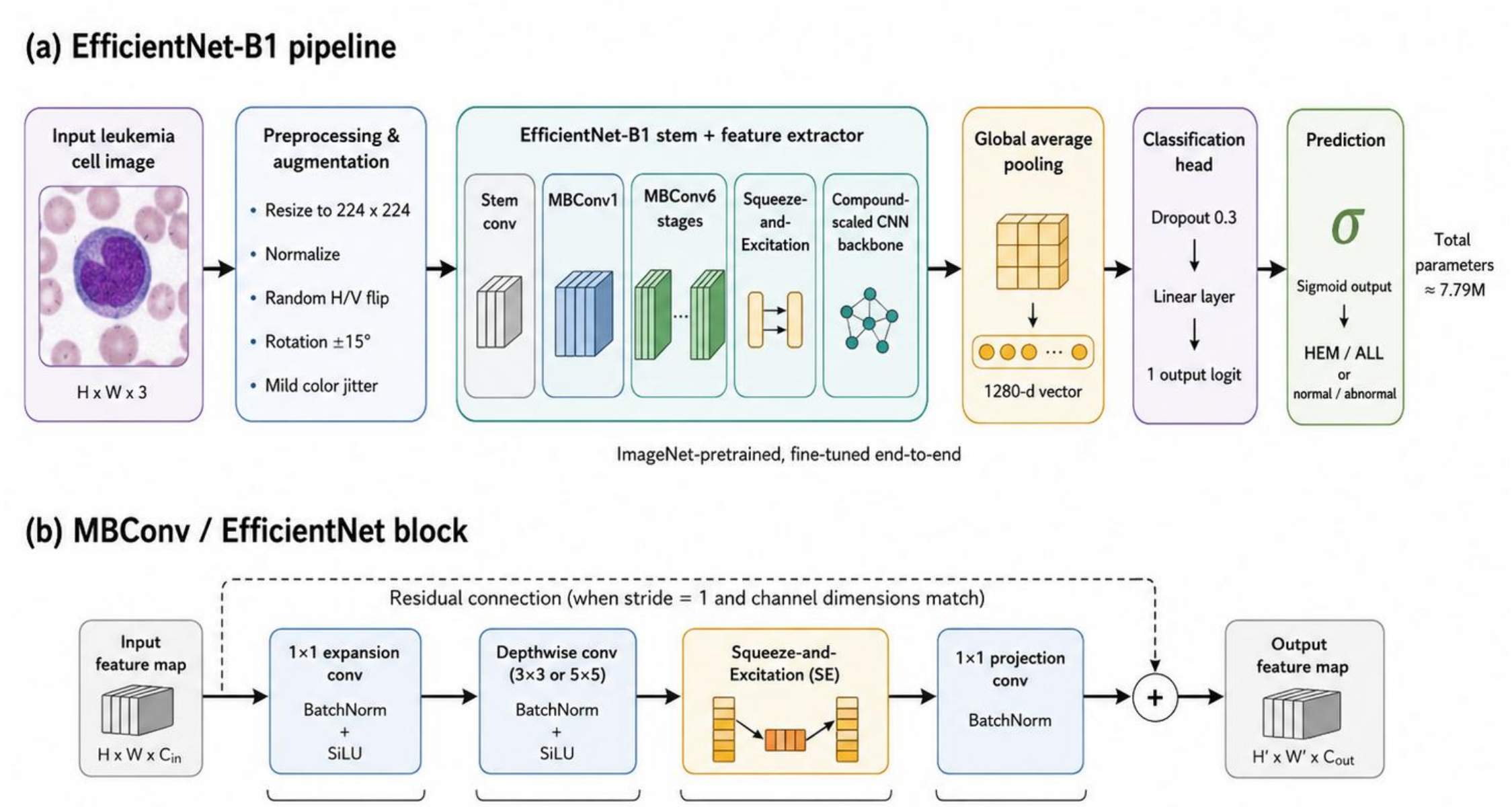

*Figure 5. Representative EfficientNet architecture used in this study (shown for EfficientNet-B1; EfficientNet-B0 shares the identical pipeline and block structure, differing only in width/depth/resolution scaling). (a) The pipeline takes a peripheral blood smear cell image, applies preprocessing and augmentation, extracts features using an ImageNet-pretrained EfficientNet backbone, aggregates them by global average pooling, and predicts the probability of ALL via a dropout-plus-linear classification head. (b) A representative MBConv block applies expansion, depthwise convolution, squeeze-and-excitation, and projection, with a residual connection when applicable.*

## 3.4 Evaluation metrics

Discrimination is assessed by accuracy, precision, recall (sensitivity), specificity, F1-score, the area under the receiver operating characteristic curve (AUROC), and the confusion matrix. Reliability is assessed by expected calibration error (ECE), Brier score, and reliability diagrams, reported before and after temperature scaling [11]. The scaling temperature is fitted on held-out cross-validation logits and then applied to the external test set, so calibration is never tuned on the test data.

The metrics are defined in terms of true positives (TP), false positives (FP), true negatives (TN), and false negatives (FN), where the malignant (ALL) class is taken as positive. Accuracy, precision, recall (sensitivity), specificity, and the F1-score are given by Equations (1)–(5):

$$\text{Accuracy} = \frac{TP+TN}{TP+TN+FP+FN} \tag{1}$$

$$\text{Precision} = \frac{TP}{TP+FP} \tag{2}$$

$$\text{Recall (Sensitivity)} = \frac{TP}{TP+FN} \tag{3}$$

$$\text{Specificity} = \frac{TN}{TN+FP} \tag{4}$$

$$F_1 = \frac{2(\text{Precision}\times\text{Recall})}{\text{Precision}+\text{Recall}} \tag{5}$$

The AUROC is the area under the curve obtained by plotting the true-positive rate (Recall) against the false-positive rate (FP / (FP + TN)) across all decision thresholds, and is equivalently the probability that a randomly chosen malignant cell receives a higher predicted score than a randomly chosen normal cell.

Calibration is quantified by the expected calibration error (ECE) and the Brier score. For ECE, predictions are partitioned into $M$ equally spaced confidence bins; the ECE is the weighted average gap between accuracy and mean confidence across bins, as in Equation (6), where $B_m$ is the set of samples in bin $m$, $n$ is the total number of samples, and acc(·) and conf(·) denote the empirical accuracy and mean predicted confidence within a bin:

$$\mathrm{ECE} = \sum_{m=1}^{M} \frac{|B_m|}{n} |\mathrm{acc}(B_m) - \mathrm{conf}(B_m)| \tag{6}$$

The Brier score is the mean squared difference between predicted probabilities and outcomes, as in Equation (7), where $p_i$ is the predicted probability of the positive class for sample $i$ and $y_i \in \{0, 1\}$ is the corresponding label:

$$\mathrm{Brier} = \frac{1}{n} \sum_{i=1}^{n} (p_i - y_i)^2 \tag{7}$$

Lower ECE and Brier values indicate better-calibrated probabilities. All calibration metrics are reported before and after temperature scaling, where a single scalar temperature is fitted on held-out cross-validation logits and applied to the external test set, so calibration parameters are never tuned on the test data.

### 3.5 Leakage ablation

A central methodological claim of this work is that the near-perfect results frequently reported on C-NMC 2019 may be inflated by patient-level data leakage rather than solely reflecting genuine diagnostic capability. To test this claim directly and to quantify the magnitude of the effect, we designed a controlled ablation that isolates the influence of the data-partitioning strategy while holding all other factors fixed.

Concretely, we compare two cross-validation schemes applied to the same pooled set of training images. In the first scheme, stratified random splitting partitions images at the image level while preserving the class ratio; because the dataset contains many cells per patient, this places cells from the same subject in both the training and validation folds and therefore permits patient-level leakage, mirroring the flawed protocol used in much of the prior literature. In the second scheme, group-wise splitting partitions images by subject, guaranteeing that all cells from a given patient remain within a single fold and thus preventing leakage. Both schemes use the same number of folds, the same features, and the same classifiers, so the only difference between them is whether patient identity is allowed to cross the train–test boundary.

We restrict this ablation to the two classical models (LightGBM and RBF-SVM) trained on frozen EfficientNet-B0 features. This choice is deliberate and yields a conservative estimate of leakage: because the feature extractor is frozen, these models cannot adapt their internal representations to patient-specific texture, staining, or acquisition cues during training. Any inflation observed under random splitting therefore arises solely from the classifier exploiting correlations between near-duplicate cells of the same patient that appear in both partitions. We hypothesize that end-to-end fine-tuned networks, which can additionally memorize subject-specific appearance, would exhibit larger leakage gaps than the frozen-feature models tested here; this expectation is not directly tested in the present study. Under that hypothesis, the value measured here would serve as a conservative lower bound on the inflation affecting the field. The resulting difference in AUROC between the two schemes, reported in Section 4.4, provides a direct, quantitative estimate of leakage-induced inflation for these two frozen-feature models.

### 3.6 Multi-seed variance protocol

Because deep-model training is stochastic, single-run point estimates can misrepresent a model's typical behavior. To characterize this variability, the three deep models (EfficientNet-B0, EfficientNet-B1, ViT-Tiny) were each trained under three random seeds (42, 123, 2024), with the identical three-fold subject-disjoint protocol, hyperparameters, and post-hoc calibration procedure repeated for every seed. Headline performance metrics for these models are reported as mean ± standard deviation across the three seeds. The frozen-feature classifiers (LightGBM, RBF-SVM) train deterministically on fixed frozen features with a fixed random state and are therefore not seed-varied; they are reported as a single run. Confusion matrices, ROC and reliability curves, Grad-CAM visualizations, and the pairwise statistical tests in Section 4.7 are computed on the canonical seed-42 run, consistent across all five models.

## 4. Results

This section reports the results of all experiments. We present external test performance across discrimination, error structure, calibration, and leakage ablation, followed by model interpretability analysis and a comparison with prior work.

### 4.1 External test performance

Table 4 summarizes performance on the external preliminary-phase test set: deep models are reported as mean ± SD across three training seeds, and frozen-feature classifiers as a single deterministic run (Section 3.6); Figure 6 visualizes the three principal metrics for the canonical seed-42 run, and Figure 7 shows the corresponding receiver operating characteristic curves. EfficientNet-B0 achieves the highest mean AUROC (0.882 ± 0.016) and the most consistent performance across seeds, ahead of EfficientNet-B1 (0.838 ± 0.028); the two EfficientNet variants clearly separate from the remaining three models. Performance decreases across families: EfficientNet-B0, EfficientNet-B1, RBF-SVM (0.781), LightGBM (0.744), and ViT-Tiny (0.640 ± 0.023). Notably, while the frozen-feature classifiers and the transformer attain high sensitivity (0.85 to 0.94), their specificity is very low (0.31 to 0.35), indicating a tendency to over-predict the malignant class; only the two EfficientNet variants exceed 0.65 specificity.

*Table 4. Performance on the external preliminary-phase test set (1,867 images, 28 unseen subjects), sorted by AUROC. Deep models (EfficientNet-B0, EfficientNet-B1, ViT-Tiny) are reported as mean ± SD across three training seeds (42, 123, 2024); frozen-feature classifiers (LightGBM, RBF-SVM) use a single deterministic run. Sens. = sensitivity/recall; Spec. = specificity; ECE (cal.) = expected calibration error after temperature scaling.*

| Model | AUROC | Acc. | Sens. | Spec. | F1 | ECE (cal.) |
|---|---|---|---|---|---|---|
| EfficientNet-B0 | 0.882 ± 0.016 | 0.818 ± 0.019 | 0.899 ± 0.028 | 0.666 ± 0.007 | 0.866 ± 0.016 | 0.033 ± 0.012 |
| EfficientNet-B1 | 0.838 ± 0.028 | 0.764 ± 0.032 | 0.813 ± 0.052 | 0.672 ± 0.009 | 0.818 ± 0.029 | 0.049 ± 0.012 |
| RBF-SVM (frozen) | 0.781 | 0.723 | 0.938 | 0.318 | 0.815 | 0.100 |
| LightGBM (frozen) | 0.744 | 0.712 | 0.929 | 0.306 | 0.808 | 0.093 |
| ViT-Tiny | 0.640 ± 0.023 | 0.678 ± 0.025 | 0.850 ± 0.061 | 0.354 ± 0.043 | 0.775 ± 0.026 | 0.098 ± 0.010 |

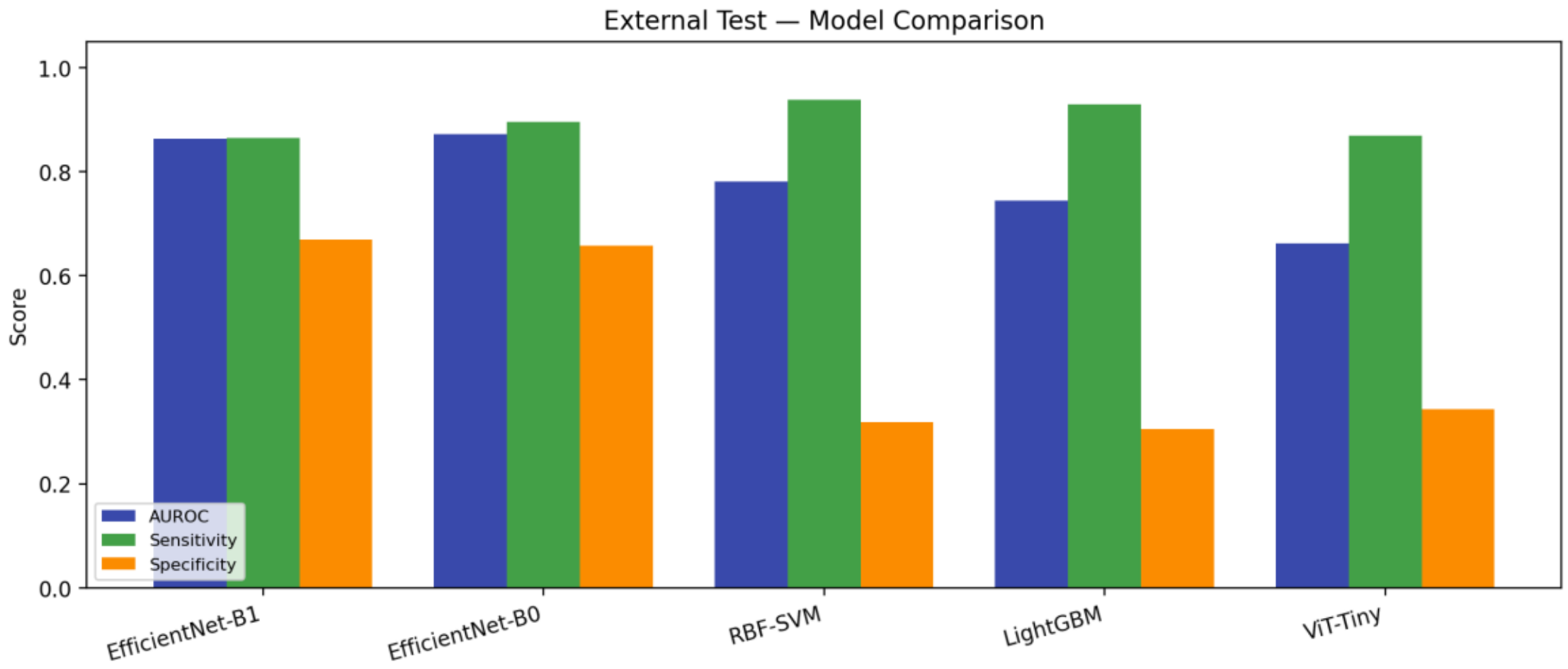

*Figure 6. AUROC, sensitivity, and specificity on the external test (canonical seed-42 run). EfficientNet-B0 and EfficientNet-B1 are the only models combining high sensitivity with high specificity; the frozen-feature classifiers and ViT-Tiny attain high sensitivity at the cost of low specificity.*

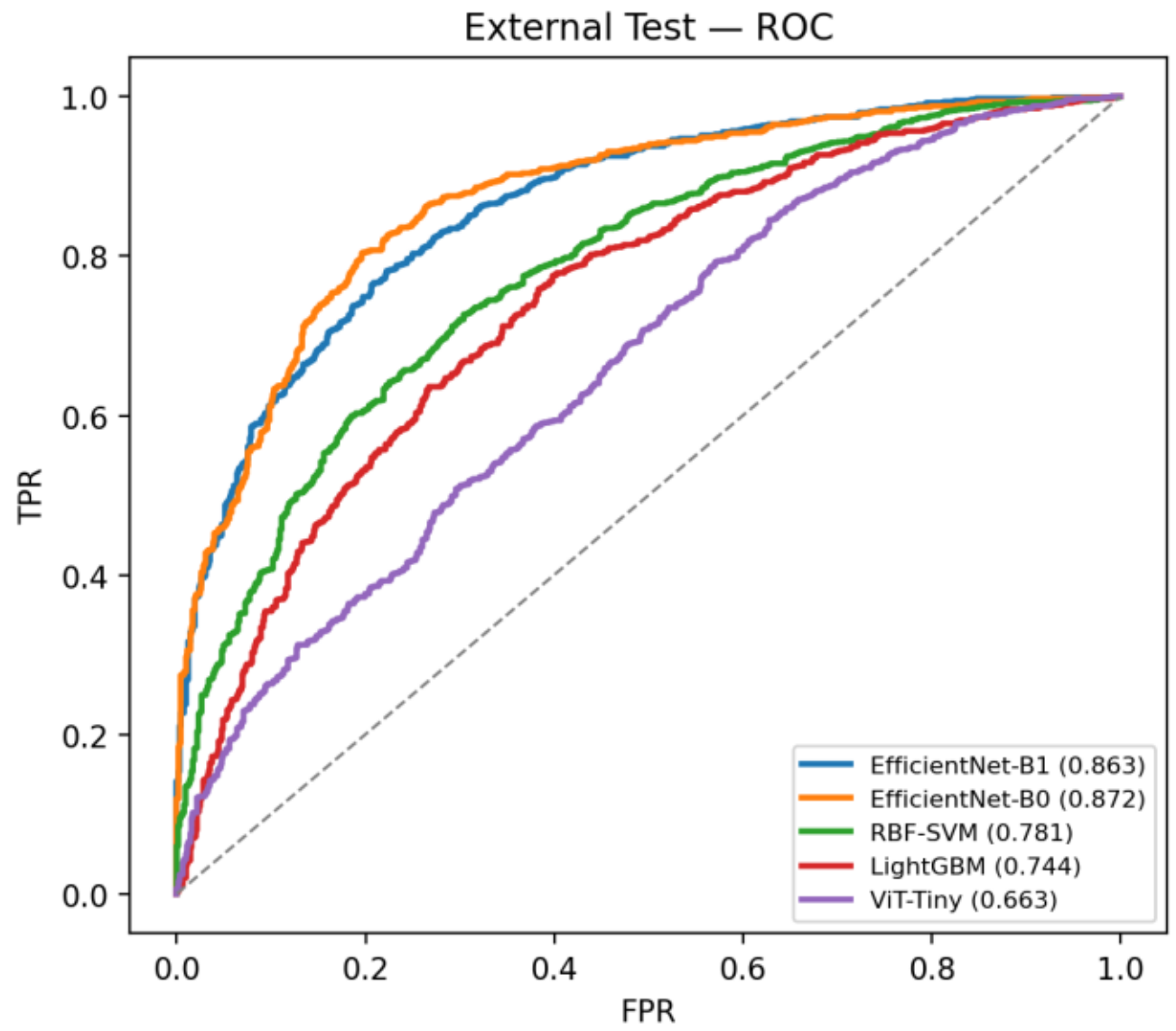

*Figure 7. Receiver operating characteristic curves on the external test (calibrated, canonical seed-42 run). EfficientNet-B0 and EfficientNet-B1 curves nearly overlap and dominate across the operating range.*

## 4.2 Error structure

The confusion matrices in Figure 8 (canonical seed-42 run) reveal a consistent failure mode for the weaker models. The frozen-feature classifiers and ViT-Tiny correctly identify almost all ALL cells but misclassify the majority of normal cells as ALL, collapsing toward the majority class. Only the two EfficientNet variants classify a clear majority of normal cells correctly—EfficientNet-B0 (426 of 648, specificity 0.66) and EfficientNet-B1 (434 of 648, specificity 0.67)—while retaining high sensitivity, indicating that sufficient model capacity, fine-tuned end-to-end, is required to learn a genuinely discriminative representation on this dataset rather than a sensitivity-maximizing shortcut.

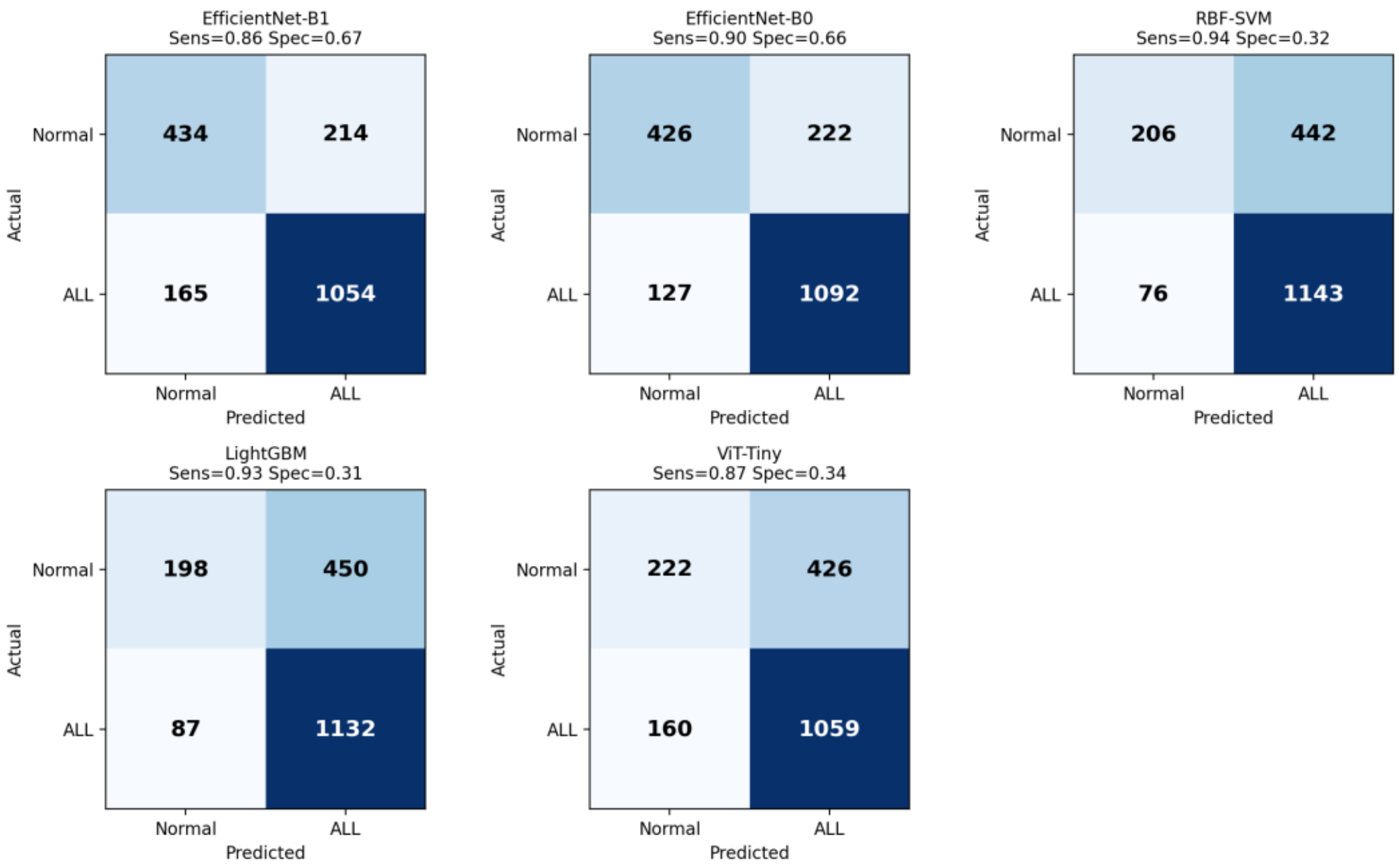

*Figure 8. Confusion matrices on the external test (calibrated, threshold 0.5). Sensitivity and specificity are annotated above each panel.*

## 4.3 Calibration

Figure 9 reports ECE before and after temperature scaling, and Figure 10 shows the corresponding reliability diagrams (canonical seed-42 run). EfficientNet-B0 is the best-calibrated model both before and after scaling (ECE 0.034 to 0.024), with EfficientNet-B1 close behind (ECE 0.082 to 0.041); the calibrated values reported in Table 4 (0.033 ± 0.012 and 0.049 ± 0.012) are the corresponding three-seed means. The frozen-feature classifiers were more overconfident and improved substantially under scaling (e.g., LightGBM 0.158 to 0.093). ViT-Tiny is the notable exception: its calibration was essentially unchanged by scaling (ECE 0.091 to 0.091, fitted temperature 1.00), indicating that the fitted temperature provided no correction on this cohort. This demonstrates that temperature scaling, while generally beneficial [11], does not guarantee improved calibration for every model, consistent with the decoupling of calibration transferability observed in cross-cohort studies [33].

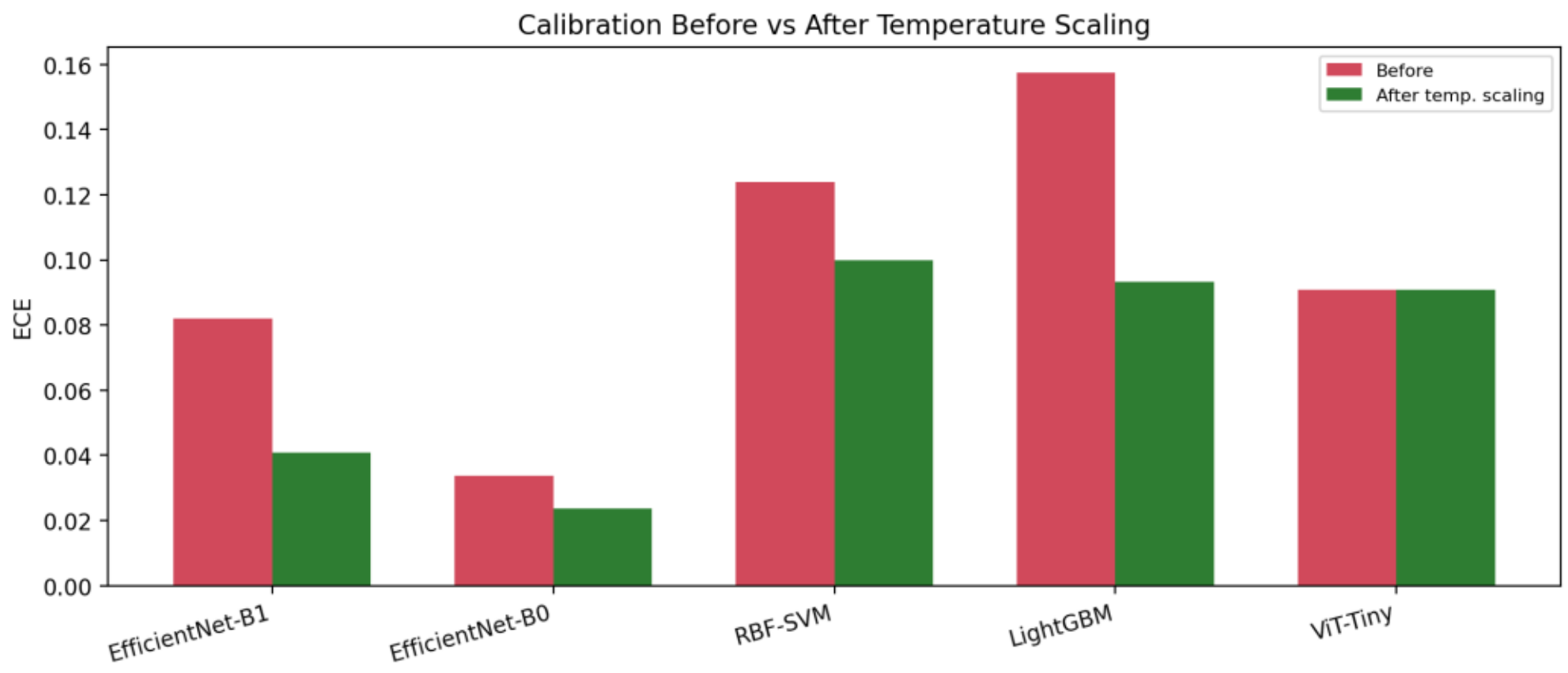

*Figure 9. Expected calibration error before and after temperature scaling. Lower is better. Scaling improves all models except ViT-Tiny, whose fitted temperature (1.00) leaves calibration unchanged.*

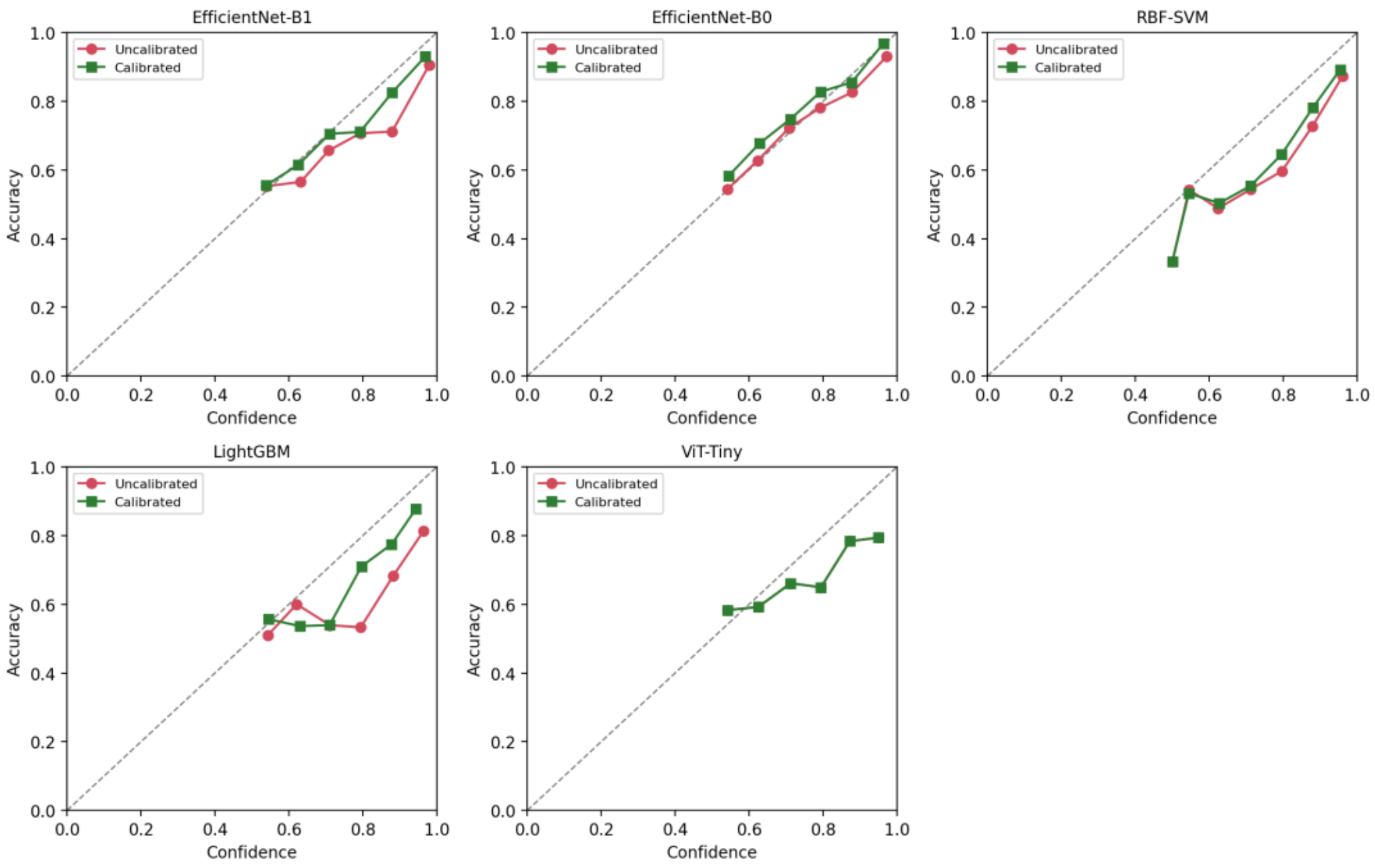

*Figure 10. Reliability diagrams (external test). Curves closer to the diagonal indicate better calibration; the calibrated EfficientNet models track the diagonal most closely.*

## 4.4 Leakage ablation

Table 5 and Figure 11 contrast random image-level splitting with subject-disjoint splitting for the two classical models on identical data. Random splitting inflates AUROC by 0.039 for both LightGBM and RBF-SVM, demonstrating measurable leakage-induced inflation for these two frozen-feature models. Because these models rely on frozen features and do not adapt to patient-specific cues during training, we interpret this as a conservative estimate; we hypothesize, but did not directly test, that end-to-end trained networks, which can memorize subject-specific texture, would exhibit larger gaps. Even for the frozen-feature models evaluated here, the effect is consistent and directionally clear: image-level evaluation overstates performance.

*Table 5. AUROC under random versus subject-disjoint cross-validation on the pooled training images. The difference quantifies leakage-induced inflation.*

| Model | Random split AUROC | Subject-disjoint AUROC | Inflation |
|---|---|---|---|
| LightGBM | 0.920 | 0.880 | +0.039 |
| RBF-SVM | 0.916 | 0.878 | +0.039 |

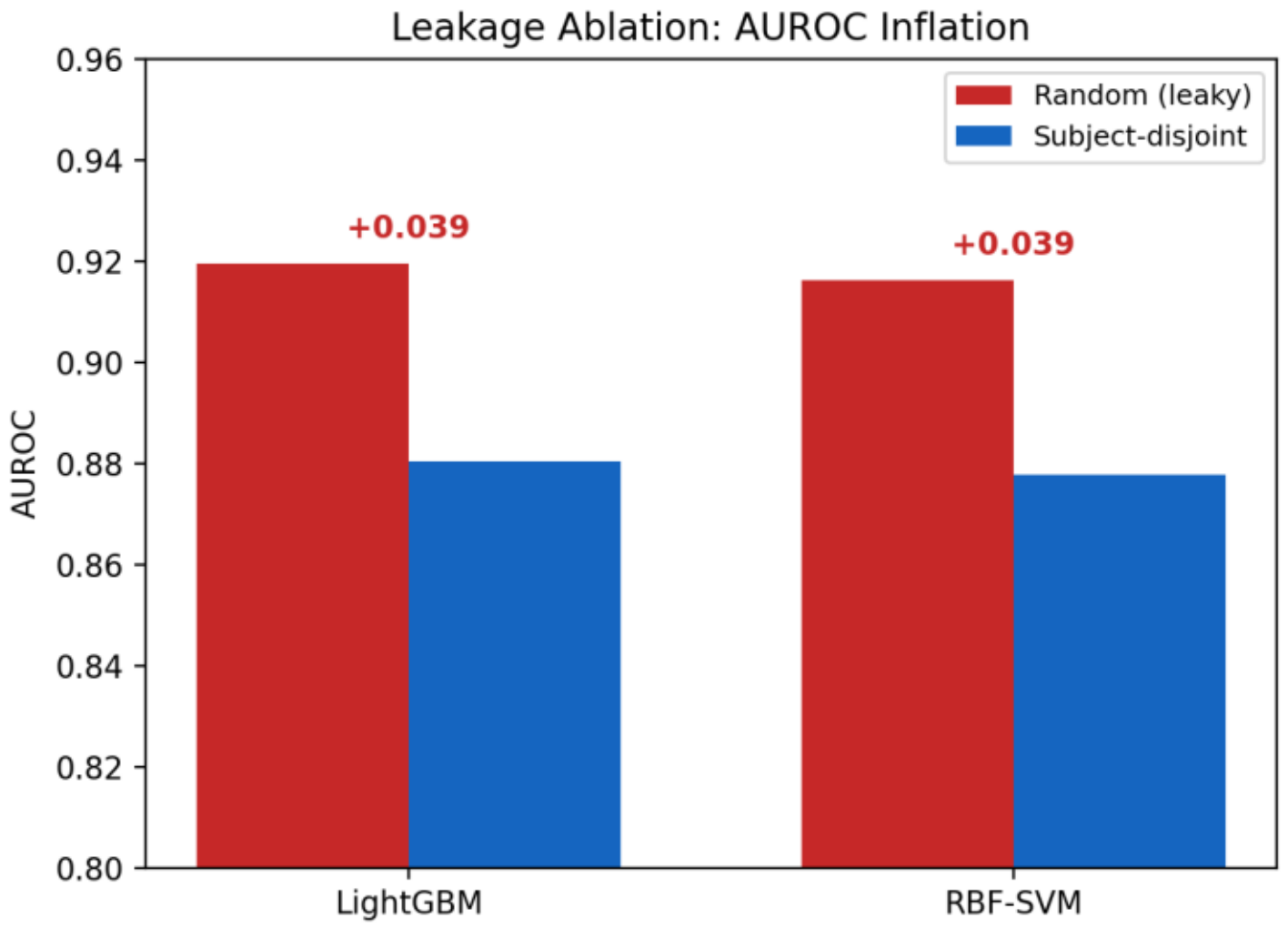

*Figure 11. Leakage ablation. Random image-level splitting inflates AUROC relative to subject-disjoint splitting for both classical models.*

## 4.5 Model interpretability

As a qualitative sanity check, Figure 12 shows Grad-CAM activation maps for representative external-test cells. Activations concentrate on the cell body and nucleus, supporting the interpretation that the network's decisions are driven by cellular morphology.

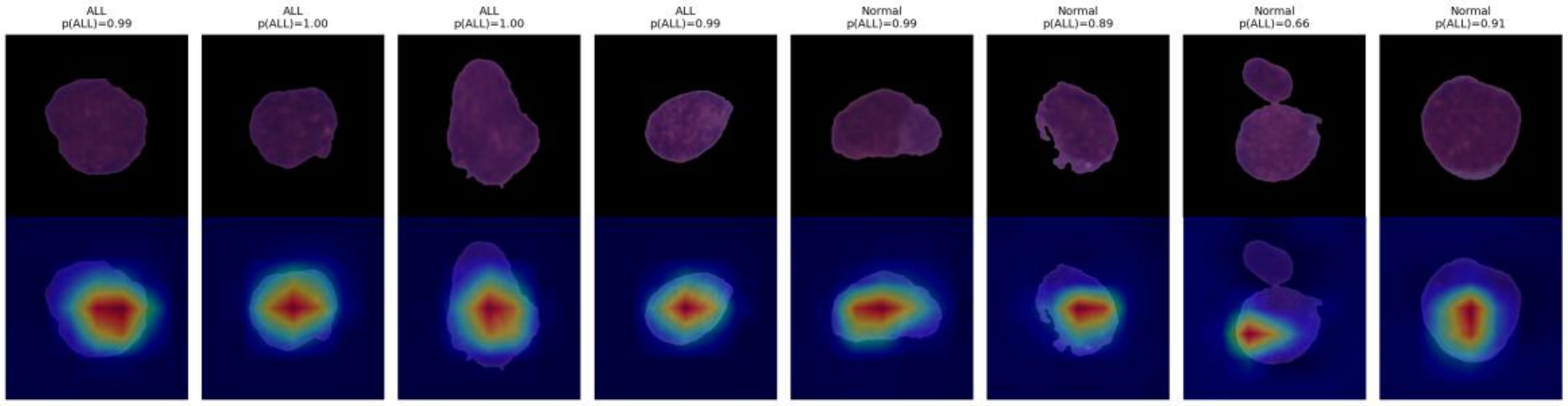

*Figure 12. Grad-CAM visualizations on representative external-test cells. Warmer regions indicate areas most influential to the malignant-class prediction.*

## 4.6 Comparison with Prior Work

Table 6 places the present results in the context of representative prior studies on C-NMC 2019. A key dimension of this comparison is the evaluation protocol: studies using image-level random splits are flagged as carrying a high risk of patient-level leakage, while those using the official challenge held-out set or a subject-disjoint protocol carry low or no leakage risk. The performance gap between prior studies—reporting accuracies and F1-scores as high as 97–98%—and the present work under subject-disjoint evaluation is consistent with the magnitude of leakage-induced inflation demonstrated for the frozen-feature models in Section 4.4. Importantly, the two studies using the official C-NMC challenge test set [2], [5] reported F1-scores in the range of 0.88–0.90, substantially lower than the image-level results, and broadly consistent with our own honestly evaluated EfficientNet-B0 result (mean AUROC 0.882 ± 0.016, F1 0.866 ± 0.016). This convergence across leakage-free evaluations reinforces the validity of the present benchmark and suggests that 0.86–0.90 F1 may better reflect the realistic difficulty of the task under leakage-free evaluation.

*Table 6. Comparison with representative prior work on C-NMC 2019. Studies using image-level random splits carry a high risk of patient-level leakage. The performance gap between prior studies (up to 97.89% accuracy / F1: 0.979) and the present work under subject-disjoint evaluation is consistent with the practical impact of data leakage on reported results, as quantified for the frozen-feature models in Section 4.4. Results marked "—" were not reported in the original study. "This work" rows (shaded blue) use subject-disjoint evaluation with no patient overlap. For the deep models, accuracy, AUROC, and F1 are reported as mean ± SD across three training seeds.*

| Study | Method | Reported Accuracy | Reported F1 / AUROC | Evaluation Protocol | Leakage Risk |
|---|---|---|---|---|---|
| Prellberg & Kramer [5] | CNN (VGG-based) | — | F1: 0.882 | Official C-NMC held-out test | Low |
| Gehlot et al. [2] | SDCT-AuxNet CNN | — | F1: ~0.90 | Official C-NMC held-out test | Low |
| Mondal et al. [4] | Weighted CNN ensemble | 86.2% | — | Image-level random split | High |
| Al-Zoubi & Al-Bzoor [3] | AutoML pipeline | ~90%+ | — | Image-level random split | High |
| Rahmani et al. [19] | ResNet + metaheuristic + MLP | 90.71% | Sens: 0.958 | Image-level random split | High |
| Awais et al. [20] | Neural ensemble + memetic opt. | ~95%+ | F1: ~0.96 | Image-level random split | High |
| Braga & Dantas [18] | EfficientNetV2-B3 + SE attention | 97.89% | F1: 0.979 | Image-level random split | High |
| **This work — EfficientNet-B0** | **Fine-tuned CNN** | **81.8 ± 1.9%** | **AUROC: 0.882 ± 0.016, F1: 0.866 ± 0.016** | **Subject-disjoint, external test** | **None** |
| **This work — EfficientNet-B1** | **Fine-tuned CNN** | **76.4 ± 3.2%** | **AUROC: 0.838 ± 0.028, F1: 0.818 ± 0.029** | **Subject-disjoint, external test** | **None** |
| **This work — RBF-SVM** | **Frozen features + SVM** | **72.3%** | **AUROC: 0.781, F1: 0.815** | **Subject-disjoint, external test** | **None** |
| **This work — LightGBM** | **Frozen features + GBM** | **71.2%** | **AUROC: 0.744, F1: 0.808** | **Subject-disjoint, external test** | **None** |
| **This work — ViT-Tiny** | **Vision transformer** | **67.8 ± 2.5%** | **AUROC: 0.640 ± 0.023, F1: 0.775 ± 0.026** | **Subject-disjoint, external test** | **None** |

## 4.7 Statistical Significance

To establish whether the performance differences in Table 4 are statistically meaningful rather than artifacts of test-set sampling, we compared the best-performing model (EfficientNet-B0) against every other model using two complementary paired tests on the external preliminary-phase set, computed on the canonical seed-42 run (Section 3.6). AUROC differences were assessed with the DeLong test for two correlated ROC curves, and differences in the correct/incorrect classification pattern at the operating threshold were assessed with McNemar's exact test. In addition, we computed 95% confidence intervals for AUROC, sensitivity, and specificity by bootstrap resampling (2,000 iterations). The DeLong test [42], McNemar's test [43], Holm correction [44], and bootstrap resampling [45] follow standard formulations.

EfficientNet-B0 significantly outperformed the two frozen-feature classifiers and the transformer on both tests, but was statistically indistinguishable from EfficientNet-B1 on this single run, as summarized in Table 7. Against EfficientNet-B1, the AUROC difference of 0.010 did not reach significance (DeLong $p = 0.150$; McNemar $p = 0.057$), and the McNemar discordance was roughly balanced—EfficientNet-B0 corrected 131 images that EfficientNet-B1 misclassified, while EfficientNet-B1 corrected 101 that EfficientNet-B0 missed. This single-run result is consistent with the multi-seed analysis in Section 3.6: EfficientNet-B0's advantage over EfficientNet-B1 is modest on the canonical seed-42 run but more consistent across seeds, with lower seed-to-seed variance (AUROC SD 0.016 vs. 0.028). The margins over the frozen-feature classifiers and the transformer were unambiguous: LightGBM (ΔAUROC 0.128, DeLong $p \approx 5.9\times10^{-25}$), RBF-SVM (ΔAUROC 0.091, DeLong $p \approx 1.2\times10^{-15}$), and ViT-Tiny (ΔAUROC 0.209, DeLong $p \approx 1.9\times10^{-71}$). Because EfficientNet-B0 was compared against four models, the p-values were interpreted after Holm–Bonferroni correction for multiple comparisons; the three comparisons against the frozen-feature classifiers and the transformer remained significant at $p < 0.001$, while the comparison against EfficientNet-B1 remained non-significant.

The bootstrap confidence intervals (patient-level, 2,000 resamples, resampling the 28 external-test patients rather than the 1,867 images) are, as expected, substantially wider than image-level resampling would suggest, and reinforce a different aspect of the picture than AUROC alone. At the patient level, the 95% AUROC interval for EfficientNet-B0 [0.767, 0.934] overlaps every other model, including ViT-Tiny [0.468, 0.797]; with only 28 independent patients, AUROC differences between models are not resolved by non-overlapping confidence intervals at this level, underscoring that the external test set, while genuinely leakage-free, offers limited patient-level statistical power. Specificity tells a clearer story: EfficientNet-B0 [0.493, 0.754] and EfficientNet-B1 [0.517, 0.773] overlap each other but neither overlaps any of the other three models (LightGBM [0.201, 0.383], RBF-SVM [0.205, 0.403], ViT-Tiny [0.151, 0.486]). This is consistent with the Discussion's central finding that specificity, not AUROC, is what reliably separates the two EfficientNet variants from the remaining models. This specificity advantage remains evident under patient-level bootstrap resampling, even though the two EfficientNet variants are not reliably distinguishable from each other on a single run. These intervals are visualized in Figure 13.

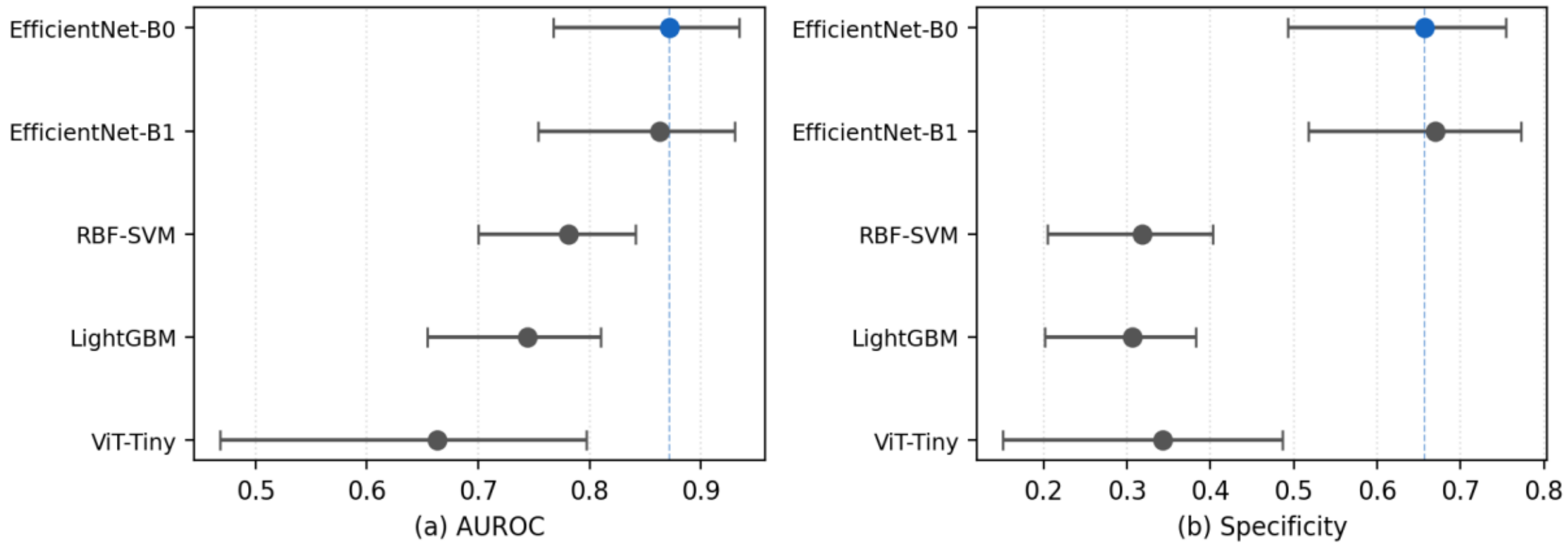

*Figure 13. Patient-level bootstrap 95% confidence intervals on the external test set for (a) AUROC and (b) specificity (canonical seed-42 run, resampled by the 28 external-test patients). AUROC intervals are wide and overlap across all five models, reflecting the limited patient-level sample size; specificity intervals for EfficientNet-B0 and EfficientNet-B1 overlap each other but neither overlaps any of the remaining three models, confirming specificity as the more robust differentiator between models at this level of analysis.*

We emphasize that the DeLong, McNemar, and bootstrap results in this section quantify uncertainty due to test-set sampling on the held-out patients, computed on a single canonical run; they do not by themselves capture variability from the stochasticity of deep-model training. That source of variability is instead characterized directly in Section 3.6 and Table 4 via repeated training under three random seeds, which is what motivates our use of the multi-seed mean as the primary reported metric in Table 4 rather than the single-run values tested here.

*Table 7. Statistical comparison of EfficientNet-B0 against each other model on the external test set (canonical seed-42 run). DeLong tests the AUROC difference; McNemar's exact test the difference in correct/incorrect classification at the operating threshold. Bootstrap 95% confidence intervals are reported in the text.*

| EfficientNet-B0 vs. | ΔAUROC | DeLong p | McNemar p |
|---|---|---|---|
| EfficientNet-B1 | +0.010 | $1.5 \times 10^{-1}$ | $5.7 \times 10^{-2}$ |
| LightGBM | +0.128 | $5.9 \times 10^{-25}$ | $1.3 \times 10^{-19}$ |
| RBF-SVM | +0.091 | $1.2 \times 10^{-15}$ | $1.0 \times 10^{-16}$ |
| ViT-Tiny | +0.209 | $1.9 \times 10^{-71}$ | $7.5 \times 10^{-33}$ |
| Comparisons against LightGBM, RBF-SVM, and ViT-Tiny are significant at $p < 0.001$ by both tests; the comparison against EfficientNet-B1 is not significant (DeLong p = 0.150, McNemar p = 0.057). ΔAUROC = AUROC(B0) − AUROC(other). | | | |

# 5. Discussion

Our results reframe what counts as strong performance on C-NMC 2019. Under subject-disjoint evaluation, no model approaches the near-perfect scores common in the literature [3]; the best mean AUROC is 0.882 ± 0.016. The gap between cross-validation and external performance, together with the leakage ablation, suggests that part of the previously reported performance may reflect evaluation protocol rather than diagnostic capability, echoing cross-cohort findings where internal performance overstates external generalization [10]. At the same time, the absence of a clear improvement from EfficientNet-B0 to the larger, more parameter-heavy EfficientNet-B1—indeed a modest average decline in mean AUROC, accompanied by a near doubling of seed-to-seed variance—suggests that architecture scaling alone does not guarantee gains on this dataset, and that additional model capacity can increase training variance without a compensating benefit in discrimination or calibration.

The single most important practical finding is that specificity, not sensitivity or AUROC alone, separates the models. Three of the five models (the two frozen-feature classifiers and the transformer) achieve high sensitivity largely by predicting ALL for most inputs, which is clinically unhelpful because it would flag most healthy cells as malignant. Only the two EfficientNet variants achieve high specificity while maintaining sensitivity; the difference between EfficientNet-B0 and EfficientNet-B1 themselves is not reliably distinguishable on a single run (Section 4.7) and is smaller, on average, than either's advantage over the remaining three models. This sensitivity–specificity tension is easily hidden by headline AUROC or accuracy and underscores the value of reporting the full confusion structure.

That the compact vision transformer ranked last, with the lowest specificity and the only calibration essentially unaffected by temperature scaling, is consistent with the view that transformers are data-hungry and underperform fine-tuned CNNs on small medical datasets [29], [30]. The calibration analysis adds a complementary dimension: a model that is well ranked may still produce untrustworthy probabilities, and consistent with reports that calibration and discrimination can transfer independently across cohorts [33], we recommend that calibration be reported alongside discrimination for any model intended for clinical decision support.

## 5.1 Limitations

The leakage ablation was conducted only for the two frozen-feature classifiers; the corresponding ablation for end-to-end trained networks, which we hypothesize may show larger inflation, is left to future work. The external test set, while genuinely subject-disjoint, originates from the same acquisition pipeline as the training data, so robustness to cross-site shift remains untested. Deep-model training remains stochastic even after our three-seed protocol (Section 3.6); a larger number of seeds, and repeated runs for the frozen-feature classifiers under alternative random states, would further sharpen the variance estimates reported here. The statistical tests in Section 4.7 are computed on a single canonical run and quantify test-set sampling uncertainty rather than training stochasticity; a fully joint treatment of both sources of uncertainty is left to future work.

A further direction is to move beyond point predictions toward selective prediction, in which the system defers its least confident cases to expert review rather than classifying every cell; such uncertainty-aware referral strategies have recently been explored for related cytology classification tasks and could extend the calibration framework established here [47].

## 6. Conclusion

We presented a leakage-aware, calibration-aware benchmark of machine learning, convolutional, and transformer models for ALL detection on C-NMC 2019 under strict subject-disjoint evaluation, incorporating multi-seed variance reporting for the deep models. Honest evaluation yields markedly lower but trustworthy performance, identifies fine-tuned EfficientNet-B0 and EfficientNet-B1 as the only diagnostically balanced models—with EfficientNet-B0 achieving the higher and more consistent mean performance across seeds, though the two were not statistically distinguishable on the canonical seed-42 run—exposes a systematic majority-class failure mode in the remaining three models, and quantifies the AUROC inflation introduced by image-level data leakage for the frozen-feature models evaluated here. We recommend that future work on C-NMC 2019 adopt subject-disjoint protocols, report calibration and full confusion structure rather than discrimination alone, and characterize training-seed variance before declaring a single architecture the winner.

## Declarations

**Funding.** This research received no specific grant from any funding agency in the public, commercial, or not-for-profit sectors.

**Conflicts of Interest.** The author declares no competing interests.

**Ethics Approval.** Not applicable. This study used the publicly available C-NMC 2019 medical imaging dataset and involved no direct interaction with human participants or collection of new patient data.

**Data Availability.** The C-NMC 2019 dataset analyzed in this study is publicly available through The Cancer Imaging Archive (TCIA) [6]. No new data were generated in the course of this work.

**Code Availability.** The code used to train, evaluate, and compare the machine learning, convolutional neural network, and Vision Transformer models reported in this study, together with the leakage-aware evaluation pipeline and related configuration files, is publicly available in the accompanying GitHub repository, https://github.com/Nisreen91/CNMC2019-Leakage-Aware-Benchmark.